\documentclass[a4paper]{jpconf}
\usepackage{graphicx}
\usepackage{amsmath,amssymb,bbm}
\begin{document}
\title{Developments in special geometry}

\author{Thomas Mohaupt${}^1$ and Owen Vaughan${}^{1,2}$}

\address{${}^1$ Department of Mathematical Sciences, University of Liverpool, 
Peach 
Street, Liverpool, L69 7ZL, UK \\
${}^2$ Department of Mathematics, University of Hamburg, Bundesstr. 55,
D-20146 Hamburg, Germany}

\ead{Thomas.Mohaupt@liv.ac.uk, Owen.Vaughan@liv.ac.uk, Owen.Vaughan@math.uni-hamburg.de}

\begin{abstract}
We review the special geometry of ${\cal N}=2$ supersymmetric
vector and hypermultiplets with emphasis on recent developments
and applications. A new
formulation of the local $c$-map based on the Hesse potential and
special real coordinates is presented. Other recent developments include the 
Euclidean version of special geometry, and generalizations of 
special geometry to non-supersymmetric
theories. As applications we disucss the proof that the
local $r$-map and $c$-map preserve geodesic completeness, and the 
construction of four- and five-dimensional static solutions 
through dimensional reduction over time. The shared features 
of the real, complex and quaternionic version of special geometry
are stressed throughout.
\end{abstract}

\section{Introduction}

The special geometry of four-dimensional vector multiplets \cite{deWit:1984pk}
has played 
a central role in studying the non-perturbative dynamics of field
theories \cite{Seiberg:1994rs,Seiberg:1994aj}, supergravity, and 
string compactifications \cite{Kachru:1995wm,Ferrara:1995yx,Kachru:1995fv}. 
It also has been
central in the studies of black holes, notably in 
the black hole attractor 
mechanism\cite{Ferrara:1995ih,Strominger:1996kf,Ferrara:1996dd,Ferrara:1996um,Behrndt:1996jn,Behrndt:1997ny,LopesCardoso:1998wt,LopesCardoso:2000qm},
and in the microscopic interpretation of 
black hole entropy in the context of string theory 
\cite{Strominger:1996sh,Maldacena:1997de}. Over time,
various re-formulations of the original definition have been found,
each with its distinguished advantages, and there has been progress 
in uncovering the underlying geometry \cite{Strominger:1990pd,Castellani:1990zd,Craps:1997gp,Freed:1997dp,1999math......1069H,Alekseevsky:1999ts}. 
By dimensional lifting and
reduction four-dimensional vector multiplets are related to 
five-dimensional vector multiplets and to three-dimensional 
hypermultiplets, respectively. The maps between the scalar geometries
induced by dimensional reduction from 5 to 4 and from 4 to 3 dimensions
are known as the $r$-map and the $c$-map, respectively 
\cite{Cecotti:1988qn,Ferrara:1989ik,deWit:1991nm,deWit:1995tf}. Since the
corresponding geometries are closely related, we will refer to all
of them as special geometries. 

While this has been an active area of research for decades, 
there are still many open questions and further directions to pursue. 
One important open problem is to master the quantum corrections in the
hypermultiplet sector of ${\cal N}=2$ string compactifications 
\cite{RoblesLlana:2006ez,Alexandrov:2008gh,RoblesLlana:2006is}. 
Here the relation between vector and hypermultiplets
induced by the $c$-map provides the starting point. Hypermultiplets are
hard to deal with because the underlying geometry, quaternion-K\"ahler
geometry, is more complicated and less understood than the other
special Riemannian holonomy geometries on Berger's list. Another developing
field is the definition and investigation of the versions of special geometry
which occur in Euclidean supersymmetric theories \cite{Cortes:2003zd,Cortes:2005uq,Cortes:2009cs}. This has two closely
related applications: the construction of instanton solutions, and,
by dimensional lifting, the construction of stationary higher-dimensional
solitonic solutions, such as black holes \cite{Mohaupt:2009iq,Mohaupt:2010du,Mohaupt:2010fk}. The Euclidean
versions of the special geometries are systematically related to
their standard counterparts through replacing complex structures by 
para-complex structures. This provides a framework for dealing with
the analytic continuations needed to describe instanton solutions 
involving axionic scalars.

The approach we are taking towards special geometry combines the
insights gained from the superconformal calculus and electric-magnetic
duality with modern differential geometry. In this article we stress
the shared features of five-dimensional and four-dimensional vector
multiplets and of hypermultiplets, and we present the corresponding
geometries as the real, complex and quaternionic version of the same
theme. In all cases, the scalar manifolds appearing in supergravity 
can be obtained by `superconformal quotients' from an associated
superconformal theory. Conversely the scalar manifolds of the 
superconformal theories are real, complex and quaternionic cones
(or at least `conical', in a sense to be made precise later)
over the scalar manifolds of the supergravity theories. 
We remark that tensor multiplets also seem to fit into this
pattern \cite{deWit:2006gn}, implying that there is yet another
type of special geometry to be placed between the complex and the
quaternionic case, since tensor multiplets have 3 real scalars. 
However, this geometry is not well understood, and we will not
discuss it in this article. 

Electric-magnetic duality occurs in four-dimensional vector multiplets
and implies the invariance of the field equations under symplectic
transformations. It also imprints itself by dimensional reduction 
on three-dimensional hypermultiplets via the $c$-map, giving the
resulting scalar manifold the structure of a symplectic vector 
bundle \cite{DeJaegher:1997ka,Cortes:2011aj}. 
One recent observation is that symplectic covariance often can be 
handled better when using a formulation of special geometry in terms
of special real instead of special holomorphic coordinates 
\cite{LopesCardoso:2006bg,Cardoso:2010gc,Cortes:2011aj}. In this
formulation the metric is a Hessian metric, i.e. it can be written
as the second derivative of a real function, the Hesse potential. 
The decisive role of such Hesse potentials is a shared feature of
all three types of special geometry discussed in this article.

One particularly interesting result is a new description of the
local (supergravity) $c$-map in terms of special real coordinates and the 
Hesse potential \cite{MohVau}.
In this formulation metrics obtained from the local $c$-map look very
similar to those obtained by the rigid version. While in \cite{Cortes:2011aj}
the terms involving the three-dimensional scalars descending from
four-dimensional gauge fields were expressed in terms of a real
coupling matrix, we have now succeded to relate this matrix to the
Hesse potential of the four-dimensional theory, and to extend this
description to all three-dimensional scalars. Since there is no
natural set of special real coordinates for the scalar manifold
of the four-dimensional supergravity theory which preserves the 
full symplectic group \cite{Ferrara:2006at}, 
we use the gauge
equivalence with the corresponding superconformal theory, for which
a description in terms of special real coordinates exists.

The formalism which we are going to review relies on the existence
of a potential, from which all couplings can be derived. It does
not require any non-generic global symmetries, and thus applies 
as well to scalar manifolds which are not homogeneous spaces (or even 
symmetric spaces). The advantages of the improved formulation and understanding
of special geometry are demonstrated by two applications. 
The first is the proof that both the $r$-map and $c$-map preserve
geodesic completeness (without any assumptions about non-generic
isometries). This is a very useful result because it provides a 
method for constructing complete, but generically non-homogeneous
quaternion-K\"ahler manifolds starting from much simpler complete
real manifolds, which are fully encoded in a homogeneous cubic 
polynomial. While the hypermultiplet manifolds of ${\cal N}=2$
string compactifications
are not geodesically complete, their finite distance singularities
are due to very specific non-perturbative effects, such as 
the appearance of additional massless modes and topological 
phase transitions. The results of \cite{Cortes:2011aj}
allow to start with a tree level
approximation which is guaranteed to be free of unphysical singularities
if the theory can be obtained from a five-dimensional theory
by dimensional reduction. 

The second application is the construction of extremal, but
not necessarily supersymmetric, multi-centered black hole
solutions in five and four dimensions by dimensional reduction 
to a Euclidean theory in four and three dimensions, 
respectively \cite{Mohaupt:2009iq,MohVau}. 
In both cases the celebrated black hole attractor equations
are derived elegantly from geometrical considerations in the
time-reduced theory. That black hole solutions are
given in terms of harmonic functions is implied by 
the field equations of the reduced theory being the 
equations for a harmonic map from space into the scalar manifold.
For BPS-type solutions, which satisfy a certain set of Bogomol'nyi type
equations, one can show both by general arguments and by the introduction
of suitable coordinates that the target of the harmonic map is a flat,
totally geodesic, totally isotropic submanifold, and, hence, 
that the non-linear second order field equations reduce to decoupled
Laplace equations.
The use of special real coordinates and
of the Hesse potential is central for directly obtaining the
manifestly symplectically covariant formulation of the attractor
equations.

Within our approach one becomes automatically aware
that there is a natural generalization
of special geometry which leads us out of the realm of supersymmetric
theories. The key features underlying the above results, namely the 
existence of a potential encoding
all couplings and the homogeneity properties implied by the relation
between superconformal and Poincar\'e supergravity, do not depend
on supersymmetry and can be generalized.
In the case of real special geometry this simply amounts to allowing 
the potential to have
an arbitrary rather than prescribed degree of homogeneity. The 
perservation of geodesic completeness by the
$r$-map, the form of the five-dimensional attractor equations, and 
the construction of five-dimensional extremal
multi-centered solutions generalize immediately 
\cite{Mohaupt:2009iq,Cortes:2011aj}. Moreover, once supersymmetry 
is not insisted on, the five dimensionally formalism can be easily be 
adapted to any dimension. However,
in four dimensions pointlike sources can carry magnetic in  addition 
to electric charge, and this implies additional features which still
need to be analysed further.  
It is not straightforward to define a generalized version of special
K\"ahler geometry, because any such generalization will imply different 
homogeneity properties for the electric and magnetic degrees of freedom.
This case requires further study.

\section{Overview of special geometries}

In this article the term `special geometry' refers to the 
geometries of vector and hypermultiplets in theories with 
eight real supercharges, corresponding to ${\cal N}=2$ extended
supersymmetry in four dimensions. Specifically, the terms `affine special 
real geometry' and `projective special real geometry' refer 
to the geometry of five-dimensional vector multiplets with rigid
and local supersymmetry, respectively, while `affine special
K\"ahler geometry' and `projective special K\"ahler geometry'
refer to four dimensional vector multiplets. The geometries 
of hypermultiplets are hyper-K\"ahler, and quaternion-K\"ahler,
respectively.

\subsection{Hypermultiplets}

Hypermultiplets exist in any dimension $d\leq 6$ and contain four
real scalars. The corresponding geometries are quaternionic. We only
consider the case where an action exists.\footnote{If only the existence
of supersymmetric equations of motion is required, the scalar manifold need
not by equipped with a metric, and the admissible geometries are more
general \cite{Bergshoeff:2002qk}.} Then the admissible scalar
geometries are Riemannian special holonomy geometries contained in 
Berger's list. For rigid hypermultiplets the geometry is 
hyper-K\"ahler \cite{AlvarezGaume:1980dk},
which means that the holonomy group of the scalar manifold $Q_{4n}$
must be contained in the compact form of the symplectic group
\[
\mbox{Hol}(Q_{4n}) \subset  USp(2n) \subset SO(4n) \;,
\]
where $n$ is the number of hypermultiples, i.e. 
$4n$ is the real dimension of $Q_{4n}$. This is equivalent to the
existence of three integrable and parallel complex structures 
$I_i$, $i=1,2,3$, which satisfy the quaternionic algebra
\begin{equation}
\label{Quat1}
I_i \circ I_j  = I_k \;,\;\;\;i,j,k \;\;\mbox{cyclic} \;,\;\;\;
(I_i)^2 = - \mathbbm{1} \;,
\end{equation}
and act isometrically. 
Therefore $Q_{4n}$ is K\"ahler with respect to all three complex 
structures.
 
If hypermultiplets are coupled to supergravity, the scalar manifold
$\bar{Q}$
must be quaternion-K\"ahler \cite{Bagger:1983tt}, which means that the holonomy group
must satisfy
\[
\mbox{Hol}(\bar{Q}_{4n}) \subset SU(2) \cdot USp(2n) \subset SO(4) \;.
\]
It is understood that the $SU(2)$ factor is non-trivial so that
hyper-K\"ahler is not a subcase. For hypermultiplets this is automatic
because of the additional condition that the Ricci scalar satisfies
\cite{Bagger:1983tt}
\[
R(\bar{Q}_{4n}) = -8n(n+1) \;.
\]
Thus supergravity hypermultiplet manifolds have a negative Ricci scalar,
and are Einstein manifolds, but not Ricci flat,
while rigid hypermultiplets manifolds are Ricci flat. Moreover 
quaternion-K\"ahler geometry is much more complicated than hyper-K\"ahler
geometry, and arguably it is the richest, most complicated and least understood
geometry on Berger's list. This is one of the main obstacles for mastering
the non-perturbative corrections to hypermultiplets arising in string
compactifications. 

Despite their name, quaternion-K\"ahler manifolds are in general neither
K\"ahler, nor even complex. The restriction of the holonomy group only
implies the local existence of three almost complex structures 
satisfying the quaternionic algebra (\ref{Quat1}). These
triplets can be patched together by $SU(2)$ transformations, but they
neither need to be globally defined, nor to be integrable. The only
global object that can be constructed is a four-form which is an 
$SU(2)$ singlet. In general one does neither have 
complex coordinates, nor a K\"ahler potential at one's disposal when
working with quaternion-K\"ahler manifolds. 

However, there is a very useful relation between quaternion-K\"ahler 
geometry and hyper-K\"ahler geometry: 
for each quaternion-K\"ahler manifold $\bar{Q}_{4n}$ of real dimension
$4n$ their exists the so-called Swann bundle or 
hyper-K\"ahler cone \cite{Swann:1991}, 
a hyper-K\"ahler manifold $Q_{4n+4}$ of real dimension $4n+4$, which 
admits a homothetic, tri-holomorphic action of the group $\mathbbm{H}^*$ 
of invertible quaternions. Conversely, every quaternion-K\"ahler manifold can be
obtained by taking a `superconformal quotient' of the associated 
hyper-K\"ahler cone:
\[
\bar{Q}_{4n} = Q_{4n+4} / \mathbbm{H}^* \;.
\]
This construction naturally occurs in the superconformal formulation 
of hypermultiplets, which uses the gauge equivalence between 
$n+1$ superconformal hypermultiplets and $n$ hypermultiplets coupled
to Poincar\'e supergravity \cite{deWit:1999fp}.\footnote{For concreteness,
we are referring to four-dimensional hypermultiplets.} 
In this context, the action of 
\[
\mathbbm{H}^* \simeq \mathbbm{R}^{>0} \cdot SU(2)
\]
corresponds to the action of dilatations and $SU(2)$ gauge 
transformations, which are part of the superconformal group,
on the superconformal hypermultiplets. The gauge-equivalent 
theory of $n$ hypermultiplets coupled to Poincar\'e supergravity
is obtained by gauge-fixing these symmetries, which geometrically
corresponds to taking a quotient with respect to $\mathbbm{H}^*$.

We will see that superconformal vector multiplets and vector multiplets
coupled to Poincar\'e supergravity are related by similar quotients,
and that the scalar manifolds of the superconformal theories are 
always cones (or at least `conical'). In the case of hypermultiplets,
the cone has four extra dimensions. While the $SU(2)$ acts isometrically,
the dilatations act homothetically, i.e. the metric of $Q_{4n+4}$ is
not invariant but rescaled by a specific constant factor. This gives
the metric of $Q_{4n+4}$ the structure of a Riemannian cone,
\[
ds^2_Q = dr^2 + r^2 ds^2_Y \;,
\]
where the $(4n+3)$-dimensional manifold $Y$ is tri-Sasakian.
$SU(2)$ acts isometrically on $Y_{4n+3}$ and  the quaternion-K\"ahler 
manifold $\bar{Q}_{4n}$ is obtained from 
$Y_{4n+3}$ by taking a quotient with respect to this action. 
We will compare this to the special geometry of vector multiplets 
in the following.

\subsection{Four-dimensional vector multiplets}

The geometries of four-dimensional vector multiplets are 
special K\"ahler geometries. They are in particular K\"ahler\footnote{Again,
we are insisting on the existence of an action principle, which implies
that the scalar manifold must carry a metric.}, and there are various
equivalent ways of defining what is `special' about them.

\subsubsection{Rigid supersymmetry}

The scalar manifolds of rigid four-dimensional vector multiplets are
affine special K\"ahler. We will discuss three definitions. 

The first definition, which is analogous to the original definition 
of projective special K\"ahler geometry  \cite{deWit:1984pk}, 
is in terms of special complex coordinates 
$X^I$, $I=1, \ldots, n$ on the scalar manifold $N_{2n}$, which 
correspond to the scalar components of vector supermultiplets
\cite{Ceresole:1994fr,Craps:1997gp}. 
A K\"ahler manifold is affine special K\"ahler
if the K\"ahler potential $K(X,\bar{X})$ can be obtained from a
holomorphic function $F(X)$, called the prepotential, by the formula
\[
K = -i (X^I \bar{F}_I - F_I \bar{X}^I)  \;,
\]
where
\[
F_I = \frac{\partial F}{\partial X^I}  \;.
\]
The origin of the existence of the additional special structure is
electric-magnetic duality. Vector multiplets contain scalars $X^I$,
fermions $\lambda_i^I$, $i=1,2$, 
and gauge fields $A^I_{\mu}$ with field strength $F^I_{\mu \nu}$.
Electric-magnetic duality is the invariance of the field equations 
(not of the action) under symplectic transformations $\Omega \in
Sp(2n,\mathbbm{R})$, which act linearly on the vector
$(F^{I}_{\mu \nu}, G_{I|\mu \nu})$ containing the field strength and dual field 
strength. The dual field strength are 
$G_{I|\mu \nu}^{\pm} \propto \partial {\cal L} / \partial F^{I|\pm}_{\mu \nu}$, 
where ${\cal L}$ is the Lagrangian, and where `$\pm$' denotes the 
projection onto the (anti-)selfdual part. Supersymmetry implies that
$(X^I, F_I)$ also transforms as a symplectic vector. 

We note in passing that there are symplectic bases where the components
$F_I$ are not the gradient of a function \cite{Ceresole:1994cx}. 
This is not a problem 
because the definition can be reformulated in terms of the 
symplectic vector $(X^I, F_I)$. Moreover, one can always go to a frame
where a prepotential exists by a symplectic transformation.

There are various ways of defining special K\"ahler geometry in 
a coordinate free way \cite{Strominger:1990pd,Castellani:1990zd,Craps:1997gp,Freed:1997dp,1999math......1069H,Alekseevsky:1999ts}. The following definition 
of affine special K\"ahler geometry is intrinsic
in the sense that it only uses the manifold $Q_{2n}$ and the canonical
bundles associated with it, i.e. the tangent and cotangent bundle 
and the resulting tensor bundles \cite{Freed:1997dp}: 
an affine special K\"ahler manifold
is a K\"ahler manifold which is equipped with a `special' connection $\nabla$ 
which is (i) flat, (ii) torsion-free,  
(iii) symplectic (the K\"ahler form is parallel), 
and satisfies (iv)
\[
d^\nabla I =0  \;,
\]
where $I$ is the complex structure (which here is interpreted as a vector 
valued one-form). 
In local coordinates 
\[
\nabla_{[a} I^b_{c]} = 0 \;.
\]
Thus the complex structure is not $\nabla$-parallel, but satisfies a weaker
condition. Note that except in the trivial case of a flat metric, the
connection $\nabla$ is different from the Levi-Civita connection $D$,
and that $\nabla$ is not metric compatible. 

For our purposes yet another definition is important, which provides a
universal construction of affine special K\"ahler manifolds in terms 
of a model vector space \cite{Alekseevsky:1999ts}: 
an affine special K\"ahler manifold is a 
complex manifold of real dimension $2n$ which locally admits a 
holomorphic and Lagrangian immersion 
\[
\Phi: Q_{2n} \rightarrow T^* \mathbbm{C}^n \simeq \mathbbm{C}^{2n} \;.
\]
Here $T^*\mathbbm{C}^n$ is interpreted as a complex-symplectic vector 
space. The special K\"ahler data on $Q_{2n}$ are induced by pulling 
back the corresponding standard data on $T^*\mathbbm{C}^n$. This definition
naturally connects to the one in terms of local coordinates. When using
standard symplectic coordinates $(X^I, W_I)$ on $T^*\mathbbm{C}^n$,
then the immersion satisfies 
\[
W_I = \frac{\partial F}{\partial X^I} \;,
\]
where the prepotential $F$ is the generating function of the 
Lagrangian immersion $\Phi =dF$. This assumes that $Q_{2n}$ is
embedded as a graph, which is the generic situation. For non-generic
immersions the function $F$ might not exist, but one can always choose
a different (generic) symplectic basis where it does.

\subsubsection{Local supersymmetry}

In the local case there exists a similar variety of equivalent defintions
of projective special K\"ahler geometry \cite{deWit:1984pk,Strominger:1990pd,Castellani:1990zd,Craps:1997gp,Freed:1997dp,1999math......1069H,Alekseevsky:1999ts}. 
None of the existing definitions
is intrinsic in the sense of the second definition of affine special 
K\"ahler geometry given in the previous section.

Like in affine special K\"ahler geometry, the K\"ahler potential
${\cal K}(z,\bar{z})$ 
of a projective special K\"ahler manifold  $\bar{Q}_{2n}$ 
can be expressed in terms of a holomorphic prepotential ${\cal F}(z)$
when using  special holomorphic coordinates $z^i$, $i=1,\ldots, n$ 
which correspond to the scalar components of vector supermultiplets: 
\begin{equation}
\label{KPproj}
{\cal K} = - \log ( -i [({\cal F} - \bar{\cal F})  - (z^i - \bar{z}^i)
({\cal F}_i + \bar{\cal F}_i)]) \;,\;\;\;
{\cal F}_i = \frac{\partial {\cal F}}{\partial z^i} \;.
\end{equation}
This formula is usually not used as a definition, but derived 
as a result.
The disadvantage of (\ref{KPproj}) is that symplectic covariance
is not manifest. The reason is that the supergravity multiplet contains
one additional vector field, the graviphoton. The total $n+1$ gauge fields and 
their duals transform linearly under the symplectic group 
$Sp(2n+2,\mathbbm{R})$, but one cannot construct a symplectic vector 
out of the $n$ scalars $z^i$. This is also a complication when 
constructing extremal black holes solutions, which work by 
`balancing' scalars against vector fields. 

There are various ways of making symplectic covariance manifest. 
We will use the original definition which arises in the context of the 
superconformal
formalism \cite{deWit:1984pk}, 
and uses the gauge equivalence between a theory of 
$n+1$ superconformal vector multiplets and $n$ vector multiples coupled
to Poincar\'e supergravity. Rigid superconformal invariance requires
that the prepotential is homogeneous of degree 2. Geometrically, this
implies 
that the scalar manifold $N_{2n+2}$ is a {\em conical} affine special
K\"ahler manifold \cite{Freed:1997dp,Alekseevsky:1999ts}, 
meaning that it admits a homothetic holomorphic
action of 
\[
\mathbbm{C}^* = \mathbbm{R}^{>0} \cdot U(1) \;.
\]
From the superconformal point of view $\mathbbm{R}^{>0}$ are the 
dilatations contained in the superconformal group, which act as
homotheties on $N_{2n+2}$, giving it the structure of a Riemannian
cone, 
\[
ds^2_N  = dr^2 + r^2 ds_S^2 \;.
\]
The basis $S_{2n+1}$ of the cone is, by definition, a Sasakian 
manifold. It can be identified with the hypersurface selected 
by the D-gauge
\[
e^{-{\cal K}} = - i (X^I\bar{F}_I - F_I \bar{X}^I) = 1 \;.
\]
The $U(1)$ transformations are likewise part of the superconformal
group and act isometrically, both on $N_{2n+2}$, and on $S_{2n+1}$.
The associated theory of $n$ vector multiplets coupled to Poincar\'e
supergravity is obtained by first gauging the superconformal transformations
and then gauge fixing the transformations not belonging to the 
Poincar\'e supergroup. The resulting scalar manifold $\bar{N}_{2n}$ 
is obtained by taking a quotient with the respect to the $\mathbbm{C}^*$
action:
\begin{equation}
\label{Kquotient}
\bar{N}_{2n} \simeq N_{2n+2}/\mathbbm{C}^* \simeq N_{2n+2}//U(1) \;.
\end{equation}
As indicated this quotient can be interpreted as a symplectic quotient
because $S_{2n+1}$ is the level set of the moment map of the $U(1)$
isometry. $\bar{N}_{2n}$ inherits a K\"ahler metric, thus the quotient
is a K\"ahler quotient. Moreover $\bar{N}_{2n}$ carries additional 
structure, and the above construction can be used as the
definition of projective special K\"ahler geometry. The relation
to the definition in terms of special coordinates is as follows: the 
homothetic $\mathbbm{C}^*$-action implies that the holomorphic
prepotential $F(X)$ of $Q_{2n+2}$ is homogeneous of degree 2, which 
is the original definition of special K\"ahler geometry. Special
coordinates $z^i$ on $\bar{Q}_{2n}$ are obtained from special 
coordinates $X^I$ on $Q_{2n+2}$ by setting $z^i=X^i/X^0$, and 
the prepotential ${\cal F}(z)$ used in (\ref{KPproj}) is related 
to $F(X)$ by
\[
F(X^0, \ldots , X^n) = (X^0)^2 F(1, X^1/X^0, \ldots ) =
(X^0)^2 {\cal F}(z) \;.
\]
It is straightforward to check that the metric induced on
$\bar{Q}_{2n}$
by the construction (\ref{Kquotient}) is a K\"ahler metric with
K\"ahler potential (\ref{KPproj}), by expressing 
\[
{\cal K} = - \log \left( -i (X^I\bar{F}_I - F_I \bar{X}^I) \right)
\]
in terms of the coordinates $X^0, z^i = \frac{X^i}{X^0}$. One then
observes that this agrees with (\ref{KPproj}), up to a K\"ahler
transformation, which removes the dependence on $X^0$.

\subsection{Five-dimensional vector multiplets}

Having reviewed the quaternionic and complex versions of special
geometry, we now turn the real case, to make some interesting observations. 
Real special geometry is 
the simplest of the special geometries, and this tends to obscure the 
analogy with the others. In the following we work out and stress
the analogies. This does not only allow us to see the systematics, but
will have important applications: the proof that the $r$-map
preserves geodesic completeness, natural generalizations of special
geometry,
and the systematic construction 
of five-dimensional black holes by lifting four-dimensional instantons.

\subsubsection{Rigid supersymmetry}

The scalar manifolds of rigid five-dimensional vector multiplets are
characterized by two properties \cite{Bergshoeff:2001hc,Cortes:2003zd}:
\begin{enumerate}
\item
The scalar metric is Hessian, i.e. it can be obtained as the second
derivative of a real function, the Hesse potential $h$:
\[
g_{ij} = \partial^2_{i,j} h \;.
\]
\item
The Hesse potential is a cubic polynomial, equivalently 
\[
\partial_k g_{ij} = \mbox{const} \;.
\]
\end{enumerate}
This definition assumes that we are using special coordinates 
$\sigma^i$, $i=1, \ldots, n$, which correspond to the scalar 
components of five-dimensional vector multiplets. 

The geometrical (coordinate free) definition of a Hessian manifold
is as follows \cite{2008arXiv0811.1658A}: 
a Hessian manifold is a (Pseudo-)Riemannian manifold
$(M,g)$,  equipped with a flat, torsion-free connection, such that the
rank 3 tensor $\nabla g$ is totally symmetric. The special coordinates
used in the first definition are the $\nabla$-affine coordinates for
which $\nabla_i = \partial_i$. In terms of special coordinates, 
the total symmetry of $\nabla g$ implies the total symmetry 
of $\partial_i g_{jk}$ (and of the Christoffel symbols of first kind),
which is the integrability condition for the existence of a Hesse
potential. 

The second condition follows from supersymmetry and gauge invariance.
Supersymmetry implies the presence of a Chern-Simons term
\[
{\cal L}_{CS} \propto C_{ijk} A^i \wedge F^j \wedge F^k \;,
\]
where $C_{ijk} \propto \partial^3_{i,j,k} h$. Gauge invariance (up to
boundary terms) requires that $C_{ijk}$ are constant. We will refer
to the scalar manifolds of rigid five-dimensional vector multiplets
as affine special real manifolds, as part of our emphasis on the 
analogies to complex special geometry. The terminology more commonly
used in the literature is `very special geometry', or `very special 
real geometry'. Affine special real manifolds are special cases
of Hessian manifolds, and it makes sense to view Hessian manifolds as a 
generalization of affine special real manifolds. For example,
the rigid version of the $r$-map (obtained by dimensional reduction 
of five-dimensional vector multiplets) has a natural generalization 
to Hessian manifolds (corresponding to the dimensional reduction 
of non-supersymmetric theories of vector and scalar fields with 
couplings encoded by a Hesse potential) \cite{2008arXiv0811.1658A}.

\subsubsection{Local supersymmetry\footnote{This section is based on 
\cite{Cortes:2011aj} and work 
in progress by the first author and Vicente Cort\'es.
}}

The coupling of five-dimensional vector multiplets and the underlying
special geometry was constructed in \cite{Gunaydin:1983bi}.
The scalar manifolds $\bar{M}_n$ of five-dimensional vector multiplets coupled
to supergravity are not Hessian, but hypersurfaces in Hessian 
manifolds. One starts with a Hessian manifold $M_{n+1}$, with
a Hesse potential $h$ that is a {\em homogeneous} cubic 
polynomial.
Then $\bar{M}_n$ is defined as a level surface of the prepotential
\begin{equation}
\label{HS}
\bar{M}_n \simeq {\cal H} = \{ h=1 \} \subset M_{n+1} \;.
\end{equation}
The natural metric on $\bar{M}_n$ is the pull back of the Hessian 
metric $g_{(0)}= -\frac{1}{3}\partial^2 h$ on $M_{n+1}$ to the 
hypersurface $\bar{M}_n$:
\begin{equation}
\label{PB1}
g_{\bar{M}} = i^* \left( - \frac{1}{3} \partial^2 h \right) \;.
\end{equation}
Here $i$ is the embedding $i: \bar{M}_n \rightarrow M_{n+1}$, and
we will explain the prefactor $-\frac{1}{3}$ below. While given
(\ref{HS}), this is the obvious choice, there are in fact 
infinitely many Hessian metrics on $M_{n+1}$ which have the same
pull back to $\bar{M}_n$. In particular, the formula given in 
\cite{Gunaydin:1983bi} is, in our conventions, 
\begin{equation}
\label{PB2}
g_{\bar{M}} = i^* \left( - \frac{1}{3} \partial^2 \log h \right) \;.
\end{equation}
It is easy to see that (\ref{PB1}), (\ref{PB2}) define the same
metric on $\bar{M}_n$ because the $\log$ only changes the behaviour
in the direction normal to the hypersurface. But as metrics on 
$M_{n+1}$ $g_{(0)} = - \frac{1}{3} \partial^2 h$ and $g_{(1)} = -\frac{1}{3}
\partial^2 \log h$ are different. 

These observations and the contrast to the quaternionic and complex
case raise the following questions. First, what are the properties of
$g_{(0)}$ and $g_{(1)}$ and what singles them out among the Hessian metrics
which have $g_{\bar{M}}$ as their pull back? Second, while $\bar{M}_n$ 
has been defined as a hypersurface, could we also regard it as 
a quotient $\bar{M} = M / \mathbbm{R}^{>0}$? Related to this, can
we regard $M$ as a cone over $\bar{M}$? 

To answer these questions we need to explore the properites of $g_{(0)}$ and 
$g_{(1)}$. We start with $g_{(0)}$ by observing that this metric is indefinite.
Imposing that the induced metric $g_{\bar{M}}$
is positive definite, as required to have standard kinetic terms for 
the scalar fields, is easily seen to imply that 
$g_{(0)}$ has Lorentz signature.
In our convention, which includes a minus sign in the definition,
$g_{(0)}$ is `mostly plus', with the negative
direction corresponding to the direction normal to the hypersurface.
The numerical factor $1/3$ has been introduced in order to comply with
conventions used in the supergravity literature. 
Next we observe that $(M,g_{(0)})$ is not a Riemannian cone over
$(\bar{M},g_{\bar{M}})$. The metric $g_{(0)}$ has a homothetic Killing vector 
\[
\xi = \sigma^I \frac{\partial}{\partial \sigma^I} \;,
\]
where $\sigma^I$ are special coordinates on $M$. Taking the Lie derivative
one finds 
\[
L_\xi g_{(0)} = 3 g_{(0)} \;,
\]
which means that the metric has weight 3 with respect to the transformation 
generated by $\xi$. Equivalently, the metric coefficients $g_{IJ}$ 
have weight 1, and are homogeneous functions of degree 1 of the special
coordinates. A Riemannian cone requires a homothety that satisfies
\[
D \xi = \mathbbm{1} \in \mbox{End}(TM)\;,
\]
where $D$ is the Levi Civita connection. Equivalently,
\[
D_X \xi = X 
\]
for all vector fields $X$. Decomposing this equation into its 
symmetric and antisymmetric part, one obtains
\[
L_\xi g = 2 g \;,\;\;\; d(g^{-1} \xi) =0 \;,
\]
where $g$ is the metric and $g^{-1}\xi$ is the one form dual to the
vector field $\xi$. In local coordinates:
\[
D_i \xi_j + D_j \xi_i = 2 g_{ij} \;,\;\;\;
\partial_i \xi_j - \partial_j \xi_i = 0\;.
\]
The first condition implies that the metric
of a Riemannian cone  
must carry weight 2, rather than 3, under the homothety, while
the second condition states that the homothetic Killing vector 
field must be hypersurface orthogonal. 

We note that for the case at hand rescaling the homothetic Killing
vector field is not an option, because we have the additional 
condition that the metric is Hessian. The homothety is the coordinate
vector field associated with the special coordinates $\sigma^I$. 
This condition can be re-expressed as
\[
\nabla \xi = \mathbbm{1}  \;,
\]
which fixes the normalization of $\xi$. The metric $g_{(0)}$
is not a Riemannian cone with respect to the homothety $\xi$ but
instead satisfies the similar condition
\[
D \xi = \frac{3}{2} \mathbbm{1} \;,\;\;\; \nabla \xi = \mathbbm{1} \;.
\]
We will call Hessian manifolds with this property 3-conical. Replacing
3 in the above definition by an arbitrary number $d$, we obtain the definition 
of a $d$-conical Hessian manifold. The case $d=2$ corresponds to 
a Riemannian cone in the usual sense. We remark that this definition 
can be adapted to K\"ahler manifolds.  The conical affine
special K\"ahler manifolds discussed before are Riemannian cones, i.e.
2-conical in the sense of the above definiton. Thus
the homothetic vector field $\xi$ which generates dilatations on conical
affine special K\"ahler manifolds satisfies 
$D\xi = \nabla \xi = \mathbbm{1}$. The $U(1)$ Killing vector feld
is given by $I\xi$, where $I$ is the complex structure.

The difference between five- and four-dimensional local vector multiplets
can be understood from the superconformal perspective. The superconformal
formulation of five dimensional supergravity has been worked out relatively 
recently \cite{Bergshoeff:2001hc,Bergshoeff:2002qk,Bergshoeff:2004kh}.
The difference between five-dimensional and four-dimensional
vector multiplets is that superconformal invariance requires 
that the four-dimensional prepotential is homogeneous of
degree 2, while the five-dimensional Hesse potential is 
homogeneous of degree 3. As a result $N$ is a cone over $\bar{N}$
while $M$, equipped with the metric $g_{(0)}$, is 3-conical. 
From the superconformal point of view it is natural to regard 
$g_{\bar{M}}$ to arise from $g_{(0)}$ by dilatational gauge fixing.
The direction normal to the hypersurface corresponds to the
compensator field for dilatational symmetry, and such fields 
typically arise with a minus sign in front of their kinetic term.

We now turn to the metric $g_{(1)}$. With the chosen sign 
this metric is positive definite. This is necessary because
the tensor field obtained by restricting $g_{(1)}$ to 
the hypersurface $h=1$, is the gauge coupling matrix of the
supergravity theory. Note that a five-dimensional supergravity 
theory with $n$ vector multiplets has $n+1$ gauge fields, because
the supergravity multiplet contains a gauge field, usually called
the graviphoton. In the superconformal approach, one starts with
a rigidly superconformal theory where both the scalar and vector
kinetic terms contain the metric $g_{(0)}$. When gauging the 
superconformal symmetries and eliminating auxiliary fields, the
gauge coupling matrix $g_{(0)}$ is replaced by $g_{(1)}$. In other
words integrating out auxiliary fields effectively replaces
the Hesse potential $h$ by its logarithm $\log h$. The metric
$g_{(1)}$ is again not a cone metric over $g_{\bar{M}}$. It is 
in fact something even simpler, namely the metric product of
a one-dimensional factor and the metric $g_{\bar{M}}$. Introducing
a coordinate $r$ by 
\[
\xi = \frac{\partial}{\partial r} = 
\sigma^I \frac{\partial}{\partial \sigma^I} \;,
\]
the metric takes the form
\[
g_{(1)} = dr^2 + g_{\bar{M}} \;.
\]
For this metric the vector field $\xi$, which acts on 
$r$ by translation and otherwise acts trivially, 
is not only a homothety but an isometry. Thus $g_{(1)}$ is
homogeneous of degree 0 in the affine coordinates $\sigma^I$,
and therefore the components $g_{(1)IJ}$ are homogenous of
degree $-2$.
We can obviously write $\bar{M}$ as a quotient, 
$\bar{M} = M / \mathbbm{R}$. To stress the analogy with the 
complex and quaternionic case, we can introduce the coordinate
$\rho = e^r$, on which $\xi$ acts by dilatations rather than 
translations, and then write $\bar{M}=M/\mathbbm{R}^{>0}$.

\section{The $r$-map}

\subsection{The rigid $r$-map}

The dimensional reduction of a theory of five-dimensional vector 
multiplets induces a map between affine special real manifolds
$M_n$ and affine special K\"ahler manifolds $N_{2n}$, called 
the $r$-map:
\[
r: M_n \mapsto N_{2n} \;.
\]
Upon dimensional reduction the components $A_5^I$ of the gauge 
fields along the reduced direction become scalars $b^I$, which
are `axions' in the sense that the five-dimensional gauge symmetry
induces an invariance under constant shifts. The resulting 
metric \cite{Cortes:2003zd,2008arXiv0811.1658A,Mohaupt:2009iq}
\begin{equation}
\label{r-map}
g_{IJ}(\sigma) d \sigma^I d\sigma^J \mapsto
g_{IJ}(\sigma) (d \sigma^I d\sigma^J + d b^I db^J)
\end{equation}
is the natural metric on the tangent bundle of $M_n$, thus
$N_{2n} \simeq TM_n$. $z^I = \sigma^I + i b^I$ are
special coordinates on $N_{2n}$, and the prepotential of $N_{2n}$ is 
related to the Hesse potential of $M_n$
by
\[
F(z^I) = h(\sigma^I + i b^I) \;.
\]
The rigid $r$-map can be generalized. If $g_{IJ}$ 
is any Hessian metric, then $g_{IJ}(\sigma) (d\sigma^I d\sigma^J + d b^I db^J)$ is
K\"ahler with K\"ahler potential
\[
K(z,\bar{z}) = h(z+\bar{z}) 
\]
and $n$ commuting isometries acting by shifts. 
Conversely, any K\"ahler metric with $n$ commuting shift isometries
can be obtained from a Hessian metric by the generalized 
$r$-map \cite{2008arXiv0811.1658A}. 

\subsection{The local $r$-map}

When coupling vector multiplets 
to supergravity the effects of dimensional reduction 
are more complicated, because the reduction of the supergravity multiplet
contributes additional degrees of freedom. The reduction of the
metric gives a vector and a scalar, and the reduction of the graviphoton
gives another scalar. 
The local 
$r$-map \cite{deWit:1991nm,Cortes:2009cs,Mohaupt:2009iq,Cortes:2011aj}
relates projective special real 
manifolds of dimension $n$ to projective special K\"ahler manifolds
of dimension $2n+2$:
\[
\bar{r} : \bar{M}_n \mapsto \bar{N}_{2n+2} \;.
\]
Since the dimension does not simply double, it is clear that 
$\bar{N}_{2n+2}$ is not the tangent bundle of $\bar{M}_n$. 
But based on the observations made above, it is nevertheless
possible to uncover the underlying geometry. $\bar{M}_n$ 
is a hypersurface given as a level set of the Hesse potential
\[
h(h^0, \ldots, h^n) = 1 \;.
\]
When performing the dimensional reduction one can 
absorbe the Kaluza-Klein scalar $\tilde{\sigma}$ into
constrained scalars $h^I$, by setting
\begin{equation}\label{RescaleReal}
\sigma^I  = e^{\tilde{\sigma}} h^I \;.
\end{equation}
The scalars $\sigma^I$ are unconstrained and can be interpreted 
as coordinats on the associated affine special real manifold
$M_{n+1}$. The metric induced on $M_{n+1}$ is the positive definite
product metric $g_{(1)} = -\frac{1}{3}\partial^2 \log h$. 
Therefore the local $r$-map can be decomposed into two operations
with a natural geometrical interpretation: the extension of $\bar{M}$
to $M$, followed by the rigid $r$-map:
\[
(\bar{M}_n, \bar{g}) \mapsto (M_{n+1}, g_{(1)}) 
\mapsto (N_{2n+2}, g_{(1)}\oplus g_{(1)}) \;,
\]
where $N_{2n+2} \simeq  TM_{n+1}$. As discussed above, $(M_{n+1}, g_{(1)})$
is a metric product, and has an isometry generated by the Killing vector field 
$\xi = \sigma^I \partial_{\sigma^I}$. This 
extends to the Killing vector field 
\[
\xi = \sigma^I \frac{\partial}{\partial \sigma^I} +
b^I \frac{\partial}{\partial b^I} 
\]
on $N_{2n}$, which combines with the $n$ shift isometries into
an $(n+1)$-dimensional solvable Lie group ${\cal L}$. This group is the
generic isometry group of projective special K\"ahler manifolds
obtained from the local $r$-map. Generic means that it only contains
the isometries generated by the $r$-map. If the manifold $\bar{M}$ 
has isometries, these will enlarge the isometry group of $N$.
As a manifold, $\bar{N}$ is locally the 
product of $\bar{M}$ and the solvable Lie group ${\cal L}$:
\[
N \simeq TM \simeq M \times \mathbbm{R}^n \simeq \bar{M} \times {\cal L}\;.
\]
This description of $N$ is crucial for proving that the local $r$-map
preserves completeness \cite{Cortes:2011aj}. 

We remark that the local $r$-map can be generalized to the case
where the Hesse potential is homogeneous of arbitrary degee $p$.
While theories with $p\not=3$ are not supersymmetric, the geometric
structure governing their bosonic sector is very similar and all
results stated above generalize in a straightforward way, because
they only depend on homogeneity, but not on the degree 
\cite{Mohaupt:2009iq,Cortes:2011aj}.

\section{The $c$-map}

We now turn to the $c$-map which is induced by the 
dimensional reduction of four-dimensional vector multiplets
\cite{Cecotti:1988qn,Ferrara:1989ik}. Our presentation focuses
on the role of the Hesse potential and of the special real
coordinates of the four-dimensional theory, which allow us
to present a new formulation of the $c$-map. In the rigid case
our re-formulation makes it manifest that the scalar manifold of the
three-dimensional is the cotangent bundle of the scalar manifold
of the four-dimensional theory, equipped with its natural metric,
while in the local case the metric is modified in a particular way.
In the local case we can also show that the manifold of the three-dimensional
theory is a group bundle, equipped with a bundle metric, which is
crucial for proving that the local $c$-map preserves completeness.

Our approach is complementary to recent constructions
of off-shell versions of the local $c$-map. The off-shell
formulation of hypermultiplets requires infinitely many
auxiliary fields. One way of dealing with this is to use
projective superspace, which is the approach taken in 
\cite{Rocek:2005ij}. The other approach is to use tensor multiplets
\cite{deWit:2006gn}. In three dimensions, hypermultiplets and
tensor multiplets are dual on-shell, but tensor multiplets admit
an off-shell formulation with finitely many auxiliary fields.
Both approaches allow to express the hypermultiplet geometry 
(specifically, the hyper-K\"ahler potential of the associated 
hyper-K\"ahler cone)
in terms of the tensor multiplet prepotential.

\subsection{The rigid $c$-map}

The dimensional reduction of four-dimensional vector multiplets
relates affine special K\"ahler manifolds $N_{2n}$ and 
hyper-K\"ahler manifolds $Q_{4n}$. This defines the rigid $c$-map
\cite{Cecotti:1988qn,Cortes:2005uq}.
Every four-dimensional gauge field
gives rise to two scalars. The first scalar is the component of the
four-dimensional gauge field along the reduced direction, 
the second scalar arises from  dualizing the three-dimensional  
vector field. The resulting supermultiplets are rigid hypermultiplets,
and so one obtains a map between affine special K\"ahler manifolds
$N_{2n}$ and hyper-K\"ahler manifolds $Q_{4n}$:
\[
c\;: N_{2n} \mapsto Q_{4n} \;.
\]
The hyper-K\"ahler manifold can be identified with the cotangent 
bundle of the special K\"ahler manifold,
\begin{equation}
\label{Cotangent}
Q_{4n} = T^* N_{2n} \;.
\end{equation}
This generalizes the statement that the cotangent bundle
of a K\"ahler manifold carries the structure of a hyper-K\"ahler
manifold in a neighbourhood of the zero 
section \cite{MR543218,1997alg.geom.10026K,Gates:1998si,Gates:1999ea}. 

The relation  (\ref{Cotangent}) 
becomes manifest when we use special real coordinates 
$(q^a)=(\mbox{Re}(z^i), \mbox{Re}(F_i))$ instead of special holomorphic 
coordinates $z^i$ on $N_{2n}$. The special real coordinates
are affine coordinates with respect to the special connection
$\nabla$, and are related to special holomorphic coordinates
by a Legendre transformation. A projective special K\"ahler 
metric is always also a Hessian metric, with the Hesse potential given by the 
Legendre transform of the imaginary part of the holomorphic
prepotential:
\[
H(q) = 2 \mbox{Im}(F)(z(q)) - 2 \mbox{Im}z^i(q) \mbox{Re}F_i \;.
\]
The rigid $c$-map takes the form
\begin{equation}
\label{RigidCmap}
g_{i\bar{j}}(z,\bar{z}) dz^i d\bar{z}^j = H_{ab}(q) 
dq^a dq^b \mapsto H_{ab}(q) dq^a dq^b + H^{ab}(q) d\hat{q}_a d\hat{q}_b\;,
\end{equation}
where $H_{ab} = \partial^2_{a,b} H$ and where $\hat{q}_a$ are the
scalars arising from dimensionally reducing and dualizing the 
gauge fields. The hyper-K\"ahler structure on $Q_{4n}\simeq T^*N_{2n}$
is given canonically in terms of the special K\"ahler data on 
$N_{2n}$ \cite{Cecotti:1988qn,Cortes:2005uq}. 
Thus like the rigid $r$-map, the rigid $c$-map has
a natural geometrical interpretation. 

\subsection{The local $c$-map}

The local $c$-map is induced by dimensionally reducing 
supergravity with $n$ vector multiplets, mapping them to 
$n+1$ hypermultiplets \cite{Cecotti:1988qn,Ferrara:1989ik}. 
The four bosonic physical degrees
of the four-dimensional gravity multiplet give rise to 
an additional hypermultiplet, usually called the universal
hypermultiplet. Thus the local $c$-map relates projective
special K\"ahler manifolds of dimension $2n$  to quaternionic 
K\"ahler manifolds of dimension $4n+4$.
\[
\bar{N}_{2n} \mapsto \bar{Q}_{4n+4} \;.
\]
The explicit form for the metric is \cite{Ferrara:1989ik,Cortes:2011aj}
\[
g_{\bar{Q}} = g_{\bar{N}} + g_G \;,
\]
where 
\[
g_{\bar{N}} = g_{i\bar{j}} dz^i d\bar{z}^{\bar{j}}
\]
is the projective special K\"ahler metric on $\bar{N}$, and
where
\begin{eqnarray}
g_G &=& \frac{1}{4\phi^2} d\phi^2 + \frac{1}{4\phi^2}
\left(d \tilde{\phi} + (\zeta^I d\tilde{\zeta}_I - \tilde{\zeta}_I 
d\zeta^I) \right)^2 + \frac{1}{2\phi} {\cal I}_{IJ}(p) d\zeta^I d\zeta^J 
\nonumber \\
&& + \frac{1}{2\phi} {\cal I}^{IJ}(p) 
(d\tilde{\zeta}_I + {\cal R}_{IK}(p) d\zeta^K)
(d\tilde{\zeta}_J + {\cal R}_{JL}(p) d\zeta^L)  \;.
\end{eqnarray}
Here $\phi$ is the Kaluza-Klein scalar, $\tilde{\phi}$ the
dualized Kaluza-Klein vector, the scalars $\zeta^I$ are the components of
four-dimensional gauge fields along the reduced direction, and
the scalars $\tilde{\zeta}_I$ are dual to the three-dimensional gauge
fields. The couplings ${\cal R}_{IJ}(p)$ and ${\cal I}_{IJ}(p)$, which 
depend on $p\in \bar{N}$ (i.e. on $z^i$), are the real and 
imaginary part of the coupling matrix of the four-dimensional 
gauge fields, 
\[
{\cal N}_{IJ} = {\cal R}_{IJ} + i {\cal I}_{IJ} 
= \bar{F}_{IJ} + i \frac{N_{IK}z^K N_{LJ} z^L}{N_{MN} z^M z^N} \;,\;\;\;
N_{IJ} = 2 \mbox{Im} F_{IJ} \;.
\]
As for the local $r$-map, the geometrical interpretation is not
immediately clear. The isometry group $G$ of $g_{\bar{Q}}$ is a 
$(2n+4)$-dimensional Lie group. The metric $g_{\bar{Q}}$ has the
structure of a `fibred product', and it was shown in \cite{Cortes:2011aj}
that the fibres, parametrized by $\phi,\tilde{\phi},\zeta^I,\tilde{\zeta}_I$
can be identified with the group $G$. Moreover, the fibres are equipped
with a $G$-invariant metric, which depends smoothly on $p\in \bar{N}$.
Thus $\bar{Q}$ is a group bundle with local form
\[
\bar{Q} \simeq \bar{N} \times G \;,
\]
and equipped with a bundle metric. One can be make this explicit by 
rewriting $g_G$ in terms of a left-invariant co-frame \cite{Cortes:2011aj}.
This rewriting of the metric is essential for proving that the
local $c$-map perserves completeness.

Another interesting way of rewriting the local $c$-map is to
use the Hesse potential (rather than the holomorphic prepotential)
of the conical affine special K\"ahler manifold $N$ associated to
the projective special K\"ahler manifold $\bar{N}$. Already in 
\cite{Cortes:2011aj} it was observed that the fibre metric $g_G$
can be written as
\[
g_G = \frac{1}{4\phi^2} d\phi^2 + \frac{1}{4{\phi}^2}
(d \tilde{\phi} + \hat{q}_a \Omega^{ab} d \hat{q}_b)^2
+ \frac{1}{2\phi} \hat{H}^{ab} d\hat{q}_a d\hat{q}_b \;.
\]
Here $(\hat{q}_a) = (\tilde{\zeta}_I, \zeta^I)$, 
\[
\Omega = (\Omega_{ab}) = \left( \begin{array}{cc}
0 & \mathbbm{1}_{n+1} \\
- \mathbbm{1}_{n+1} & 0 \\
\end{array} \right) \;,\;\;\;
\Omega^{-1} = (\Omega^{ab}) = \left( \begin{array}{cc}
0 & -\mathbbm{1}_{n+1} \\
+ \mathbbm{1}_{n+1} & 0 \\
\end{array} \right) \;,\;\;\;
\]
and 
\[
\hat{H} = (\hat{H}_{ab}) = \left( \begin{array}{cc}
{\cal I} + {\cal R} {\cal I}^{-1} {\cal R} & - {\cal R} {\cal I}^{-1} \\
- {\cal I}^{-1} {\cal R} & {\cal I}^{-1} \\
\end{array} \right) \;,\;\;\;
\hat{H}^{-1} = (\hat{H}^{ab}) = \left( \begin{array}{cc}
{\cal I}^{-1} & {\cal I}^{-1} {\cal R} \\
{\cal R}{\cal I}^{-1} & {\cal I} + {\cal R}{\cal I}^{-1} {\cal R}
\\ 
\end{array}\right) \;.
\]
Note that $\hat{q}_a$, $\hat{H}_{ab}$ and $\hat{H}^{ab}$ transform
linearly under symplectic transformations, whereas ${\cal N}_{IJ}$
transforms fractionally linearly. In the above expression for $g_G$
symplectic invariance is manifest. 

More recently, this rewriting has been extended to the remaining
variables $z^i, \phi, \tilde{\phi}$, resulting in an expression 
which is very similar to the `metric on the (co)tangent bundle form'
of the rigid $c$-map (\ref{RigidCmap}) \cite{MohVau}. 
This requires the following
series of observations. First, the Hesse potential is associated to 
the conical affine special K\"ahler manifold $N$, and symplectic 
transformations act in a simple way on $N$, but not on $\bar{N}$. 
Therefore it is not possible to introduce special real coordinates
on $\bar{N}$ which transform linearly under the full symplectic
group \cite{Ferrara:2006at}. To circumvent this problem 
we re-express $g_{\bar{N}}$ in terms of quantities defined on $N$ by
\[
g_G = g_{i\bar{j}} dz^i d\bar{z}^{\bar{j}} = 
g_{IJ} dX^I d\bar{X}^J \;,
\]
where
\[
g_{IJ} = \frac{\partial^2 {\cal K}}{\partial X^I \partial \bar{X}^J} \;,\;\;\;
{\cal K} = - \log \left(- i (X^I \bar{F}_I - F_I \bar{X}^I)\right) \;,
\]
is a degenerate tensor field on $N$ which by pull back gives the
(non-degenerate) metric on $\bar{N}$. The expression 
$g_{IJ} dX^I d\bar{X}^J$ formally depends on two additional degrees 
of freedom, corresponding to the radial direction of the cone $N$ over
the Sasakian $S$, and to the orbits of the $U(1)$ action on $N$. The first
is eliminated by imposing the D-gauge\footnote{We remark that this degree of
freedom can be isolated and decoupled by suitable field redefinitions.}
\[
-i (X^I \bar{F}_I - F_I \bar{X}^I ) = 1 \;.
\]
Moreover $g_{IJ} dX^I d\bar{X}^J$ is invariant under local $U(1)$ 
transformations, which eliminate the second additional degree of freedom.

The next step is to absorbe the Kaluza-Klein scalar into 
the fields $X^I$ living on $N$, as we did for the local
$r$-map in (\ref{RescaleReal}).
Defining\footnote{When comparing to \cite{MohVau}, one has to
replace $\phi \rightarrow e^\phi$. }  
\[
Y^I = \phi^{1/2} X^I \;,
\]
the Kaluza-Klein scalar becomes a dependent field:
\[
\phi = - N_{IJ} Y^I \bar{Y}^J = -i (Y^I \bar{F}_I - F_I \bar{Y}^I)\;.
\]
This identifies the Kaluza-Klein scalar with the radial direction
on $N$, which now has become a dynamical degree of freedom. Since
the local $U(1)$ invariance is intact, the $Y^I$ correspond to 
$2n+2-1=2n+1$ real scalars. Together with the dualized Kaluza-Klein
vector $\tilde{\phi}$, which remains an independent field, 
we have $2n+2$ real scalars, while the scalars $\hat{q}_a$ 
obtained from the four-dimensional gauge fields add another 
$2n+2$, bringing the count to $4n+4$. 

The next step is to replace the $Y^I$ by the corresponding 
special real coordinates $q^a$. Like the $Y^I$, the $q^a$ are
subject to a local $U(1)$ transformation and therefore only
correspond to $2n+1$ independent real scalars. Using that the
$q^a$ are special real coordinates on $N$, one can show that
\[
g_{IJ} dX^I d\bar{X}^J = \left( - \frac{1}{2H} H_{ab} + \frac{1}{4H^2}
H_a H_b + \frac{1}{H^2} (\Omega_{ac} q^c \Omega_{bd} q^d)
\right)dq^a dq^b\;,
\]
where $H_a$ are the first derivatives of the Hesse potential.
Next we remember that in the five-dimensional case the metric
obtained after absorbing the Kaluza-Klein scalar could be 
expressed in terms of the logarithm of the Hesse potential of
the rigid theory. This motivates us to define
\[
\tilde{H} = - \frac{1}{2} \log H \;,\;\;\;
\tilde{H}_{ab} = \partial^2_{a,b} \tilde{H} \;,
\]
and we find
\[
g_{IJ} dX^I d\bar{X}^J = \left(\tilde{H}_{ab} - \frac{1}{4H^2}
H_a H_b + \frac{1}{H^2} (\Omega_{ac} q^c \Omega_{bd} q^d)
\right) dq^a dq^b \;.
\]
It turns out that the term proportional to $H_a H_b$ now cancels against
the term $(4\phi)^{-2} d\phi^2$ in $g_G$. To complete the rewriting,
we remark that while the symplectic tensor $\hat{H}_{ab}$, which encodes
the four-dimensional gauge couplings, is not a Hessian metric, it is
related to the Hessian metric $\tilde{H}_{ab}$ by
\[
\tilde{H}_{ab} = \frac{1}{H} \hat{H}_{ab} - \frac{2}{H^2}
(\Omega_{ac} q^c \Omega_{bd} q^d) \;.
\]
This allows us to rewrite the metric $g_{\bar{Q}}$ such that the
dependence on the underlying Hesse potential is exclusively 
through the Hessian metric $\tilde{H}_{ab}$, and $H$ itself
(which is proportional to the Kaluza-Klein scalar):\footnote{
For convenience, we have replaced the coordinates $\hat{q}_a$ by 
their duals $\hat{q}^a$, defined by 
$d\hat{q}^a = H^{ab}d\hat{q}_b$. } 
\begin{equation}\label{MetricQK}
g_{\bar{Q}} = \tilde{H}_{ab} (dq^a dq^b +
d\bar{q}^a d\bar{q}^b ) 
+ \frac{1}{H^2} (q^a \Omega_{ab} dq^b)^2 
+ \frac{2}{H^2} (q^a \Omega_{ab} d\hat{q}^b)^2 
+ \frac{1}{4H^2} ( d \tilde{\phi} + 2 \hat{q}^a \Omega_{ab}
d\hat{q}^b)^2 \;.
\end{equation}
This form of the metric is very close indeed to the form of
metrics obtained from the rigid $c$-map, with the couplings of the
additional terms only depending on the constant matrix $\Omega$ and the 
Kaluza-Klein scalar.


\section{Geodesic completeness}

The local $r$-map allows to generate special K\"ahler manifolds from 
simpler special real manifolds, while the $c$-map generates 
quaternion-K\"ahler manifolds form special K\"ahler manifolds. 
By combining both maps one can start with a relatively simple
special real manifold, encoded in a homogeneous cubic polynomial
and construct an associated quaternion-K\"ahler manifold. 
This is very useful, because a complicated problem is related
to a simpler one, both mathematically, describing quaternion-K\"ahler
manifolds in terms of special real manifolds, and physically, 
describing hypermultiplets in terms of vector multiplet data.
One particularly
interesting feature of this construction is that when starting 
with a non-homogeneous special real manifold, the result will be
a non-homogeneous quaternion-K\"ahler manifold, at least generically.\footnote{
There is no theorem forbidding that the result is homogeneous or 
symmetric, but the counting of the generic isometries generated by
the $r$- and $c$-map indicates that such cases are exceptional.}
Thus the combined $r$- and $c$-map is a tool to construct non-homogeneous
quaternion K\"ahler manifolds with all data encoded in a homogeneous
cubic polynomial. One then naturally wonders about the properties
of the manifolds produced in this way. Geodesic completeness is
a very important geometrical property, which in physical terms means 
that no singularity
in coupling space can be reached in finite time. 

In the past, the symmetric and homogeneous 
spaces generated by the $r$-map and $c$-map have been
studied exhaustively, including the classification of
the homogeneous quaternion-K\"ahler
spaces arising from the $r$ and $c$-map \cite{MR0402649,deWit:1991nm,MR1395026}.
Homogeneous spaces are geodesically complete, and it is 
known that the $r$-map and $c$-map relate homogeneous spaces to 
homogeneous spaces. 

The improved understanding of the geometry of the local $r$- and
$c$-map allows us to prove a very interesting statement which 
extends the classical results reviewed above \cite{Cortes:2011aj}: 
the generalized\footnote{Here we refer to the version of the local $r$-map
where the degree of homogeneity of the Hesse potential is abitrary.}
local $r$-map and the local $c$-map preserve geodesic completeness.
Thus complete special real manifolds can be used to construct 
complete, but generically non-homogeneous quaternion-K\"ahler 
manifolds. Moreover, the classification of complete special real
manifolds appears to be tractable, at least in low 
dimensions \cite{Cortes:2011aj}.

The preservation of completeness follows from 
a general theorem, which states that given a complete 
Riemannian manifold, the Riemannian metrics on certain bundles
are complete as well \cite{Cortes:2011aj}. Specifically, given a complete 
Riemannian manifold $(M_1,g_1)$ and a smooth family 
$g_2(p)$, $p\in M_1$ of $G$-invariant metrics on a homogeneous
manifold $M_2=G/K$, then the metric $g=g_1+g_2$ on $M_1\times M_2$
is complete, with isometric action of $G$. Moreover, this 
result generalizes from global products to bundles which take
this form in a local trivilization. Since manifolds in the image
of the local $r$-map and local $c$-map have the required form, it 
follows that both maps preserve completeness.

Let us indicate how the theorem is proved. Remember that
there are two relevant concepts of completeness for Riemannian
manifolds. Metric completeness means that all Cauchy series 
converge, geodesic completeness means that every geodesic ray
can be extended to infinite length. The Hopf-Rinow theorem 
states that both conditions are equivalent to one another.
One can show that completeness is also equivalent to the condition 
that every curve which is not contained in any compact subset 
has infinite length. Then the theorem is proved by estimating
curve lengths on $M_1\times M_2$  using that $M_1$ is assumed 
to be complete. For the problem of classification of complete special real
manifolds  we refer to \cite{Cortes:2011aj} and future work.

While the mathematical merits of this result
are obvious, the physical implications require further comment.
If one was to consider supergravity as a fundamental theory,
one would require that the coupling space has no singularities
at finite distance, and therefore impose that the scalar manifold
must be geodesically complete. However, the more likely candidate
for a fundamental theory is string theory, with supergravity as 
a low-energy effective description. The scalar manifolds arising
in ${\cal N}=2$ supersymmetric string compactifications are not
complete, but rather `incomplete in an interesting fashion.' 
More precisely, one expects and indeed finds singularities at finite distance
which corresponding to special loci in the moduli space where 
additional massless states occur. The most prominent example
is the conifold singularity of Calabi-Yau threefolds \cite{Strominger:1995cz}. 

Despite that string moduli spaces are not complete, we expect
that our result is a significant step towards understanding
the global geometry of string moduli spaces, in particular
hypermultiplet moduli spaces. The incorporation of perturbative
and non-perturbative corrections to the $r$- and $c$-map,
and the study of their role in the global geometry of
string moduli spaces is an interesting topic for future
work.

\section{Generating stationary solutions}

Dimensional reduction over time is a method which allows to construct
stationary solutions. If all relevant fields in the reduced theory
are scalars, the remaining problem of solving the scalar field
equation is equivalent to constructing a harmonic map from 
space into the scalar manifold \cite{Breitenlohner:1987dg}. 
One standard approach is to
find totally geodesic submanifolds and to find harmonic maps from 
space-time  into them. An important subclass of solutions, which lifts
to (supersymmetric as well as non-supersymmetric) 
extremal black holes, is provided by maps into submanifolds
which are totally isotropic in addition to being totally geodesic.

In this section we will review how the temporal versions of the $r$-map and
of the $c$-map can be used in the context of this 
construction \cite{Cortes:2009cs,Mohaupt:2009iq,MohVau}. 
In particular, we will see that a whole class of totally geodesic, 
totally isotropic submanifolds can be identified. Moreover, there is a 
set of canonical coordinates on these submanifolds which reduce
the non-linear second order scalar field equations (corresponding to a harmonic
map between Riemannian manifolds) to decoupled linear harmonic
equations. 
We will also see that the five-dimensional and four-dimensional
black hole attractor equations can be derived from Bogomol'nyi 
equations of the time-reduced theory.

Before starting, let us comment on the relation between our 
approach and others. Originally, the black hole attractor or
stabilization equations were derived by imposing supersymmetry,
i.e. the existence of Killing spinors. Imposing that the event
horizon is finite implies enhanced supersymmetry on the horizon and
forces the scalar fields to take prescribed fixed point values which
are completely determined by the 
charges \cite{Ferrara:1995ih,Strominger:1996kf,Ferrara:1996dd,Ferrara:1996um}. The attractor equations 
determining the fixed point values can be formulated as a symplectically
covariant equation relating two symplectic vectors, one containing
the electric and magnetic charges, the other being proportional to 
the imaginary part of $(X^I,F_I)$, schematically
\begin{equation}
\label{Attr1}
(\mbox{Im} X^I, \mbox{Im} F_I ) \sim (p^I, q_I) \;.
\end{equation}
This equation admits a generalization
which specifies the black hole solutions globally (not only at the
horizon) in terms of harmonic functions. As before there is an 
equation between two symplectic vectors, one containing the harmonic
functions, the other is again proportional to the imaginary 
part of $(X^I, F_I)$. Schematically
\begin{equation}
\label{Attr2}
(\mbox{Im} X^I, \mbox{Im} F_I ) \sim (H^I, H_I) \;.
\end{equation}
From these generalized attractor equations
(also called generalized stabilization equations), the previous
equations can be recovered by taking the near horizon limit. 
It was shown that the relations 
(\ref{Attr2}) are not only sufficient \cite{Behrndt:1997ny}, 
but 
also necessary to obtain supersymmetric 
solutions \cite{LopesCardoso:2000qm}. Moreover,
it is possible to include a class of higher-derivative terms
\cite{LopesCardoso:2000qm}.

It was already observed in \cite{Ferrara:1997tw} that the attractor equations
could also be obtained from the equations of motion. Solving the scalar 
equations
of motion can be re-formulated as a problem involving geodesic
motion, and it is natural to combine this with dimensional reduction.
Starting from \cite{Goldstein:2005hq,Tripathy:2005qp,Kallosh:2005ax}
this approach has been used to construct non-supersymmetric extremal
solutions. The supersymmetric attractor equations can be formulated
as gradient flow equations, which are driven by the central 
charge \cite{Ferrara:1997tw,Moore:1998pn,Denef:2000nb}. For non-supersymmetric
solutions it is also possible to reduce the second order field equations
to first order gradient flow equations, which are then driven by a 
different function, often called a fake superpotential \cite{Ceresole:2007wx,LopesCardoso:2007ky,Perz:2008kh,Ortin:2011vm}. 
The flow equation  involve the physical scalars
$z^i$ rather than the symplectic vector $(X^I,F_I)$ and are not
manifestly symplectically covariant. For symmetric scalar target spaces,
the construction of both supersymmetric and non-supersymmetric 
extremal solutions can be related to integrable systems 
\cite{Gunaydin:2005mx,Gunaydin:2007bg,Bossard:2009we},
\cite{Chemissany:2009hq,Andrianopoli:2009je,Chemissany:2010zp}.

Our approach is somewhat different in that while we also impose
Bogomol'nyi equations, we do not derive gradient flow equations.
Instead we solve the second order field equations directly in 
terms of harmonic functions and obtain the solution as an equation
between two symplectic vectors, one containing the harmonic functions,
the other encoding the scalars. Our method does not rely
on non-generic isometries (as those needed to have a symmetric target
space), but on the existence of a Hesse potential or holomorphic
prepotential which encodes the geometry. Using the 
para-complex geometry of the target spaces obtained by dimensional
reduction over time allows to identify totally geodesic, totally
isotropic submanifolds which correspond to solutions. Special geometry
also provides adapted coordinates which allow to reduce
the non-linear second order equations to decoupled harmonic equations.
Obtaining multicentered static extremal solution is as easy as
obtaining single-centered static extremal solutions. The relation 
of our approach to the one based on gradient flow equations and
integrability is not completely understood, but some aspects have
been discussed in \cite{Mohaupt:2009iq}.

We remark that while we are focusing on using Euclidean 
solutions as a tool for generating black hole solutions,
our Euclidean solutions are valid instanton solutions which
are interesting in their own right \cite{Mohaupt:2010du}.
This is an interesting topic in itself, see for example 
\cite{AzregAinou:2011gt} for recent work on the classification
of instantons in Einstein-Maxwell type theories.

In the above discussion we have referred to the four-dimensional
version of the attractor mechanism for concreteness. There is
also a five-dimensional 
version \cite{Chamseddine:1996pi,Chamseddine:1998yv,Chamseddine:1999qs}
which is accessible to our method. We will discuss this case 
first, as it is technically simpler. 

\subsection{Generating solutions from the local $r$-map}

It is straightforward to generalize both the rigid and the
local $r$-map such that one treats dimensional reduction 
over space, $\epsilon=-1$, and dimensional reduction over
time, $\epsilon =1$, in parallel:
\begin{equation}
\label{r-map_epsilon}
g_{IJ}(\sigma) d \sigma^I d\sigma^J \rightarrow
g_{IJ}(\sigma) (d \sigma^I d\sigma^J - \epsilon d b^I db^J) \;,
\end{equation}
where 
\[
g_{IJ} \sim - \partial^2_{I,J} h 
\]
in the rigid case and
\[
g_{IJ} \sim - \partial^2_{I,J} \log h 
\]
in the local case. For time-like reductions the scalars 
coming from higher-dimensional gauge fields enter with the
opposite sign. This implies that the scalar manifold $N_{2n}$
has indefinite (`split') signature $(+)^n(-)^n$, and therefore
has totally isotropic submanifolds of dimension $n$.

While for space-like reductions the scalars
naturally combine into complex scalars $X^I = \sigma^I + i b^I$,
for time-like reductions they combine into para-complex
scalars
\[
X^I = \sigma^I + e b^I \;,
\]
where $e^2=1$, $\bar{e}=-e$. Para-complex geometry is in many
respects analogue to complex geometry, and one can view complex
and para-complex geometry as different real forms of 
complex-Riemannian geometry.
The concepts of para-Hermitian, 
para-K\"ahler, special
para-K\"ahler, para-hyper-K\"ahler and para-quaternion-K\"ahler
manifolds can be defined, and play a role in the Euclidean
version of special geometry \cite{Cortes:2003zd,Cortes:2005uq,Cortes:2009cs}. 
In this article we will not make
heavy use of this formalism. In particular, we will not 
use para-holomorphic coordinates, but focus on suitable 
real coordinate systems instead. 

At this point we can introduce another generalization. 
Supersymmetry requires that the Hesse potential has degree
3 (rigid case) or even is 
homogeneous of degree 3 (local case). However the rigid $r$-map
works for any Hessian manifold \cite{2008arXiv0811.1658A}, and the local $r$-map
can be generalized to the case where $h$ is a homogeneous
function of arbitrary degree $p$ \cite{Mohaupt:2009iq,Cortes:2011aj}. 
We will refer to this
class of manifolds as generalized special real geometry.
For $p\not=3$ there are
no associated supersymmetric theories, but there is a 
corresponding class of non-supersymmetric theories 
of scalars, gauge fields and gravity, with the couplings
controled by the generalized special real geometry. 
The Lagrangian takes the form
\[
{\cal L} \sim - \frac{1}{2} R_{(5)} - g_{IJ} \partial_\mu h^I \partial^\mu 
h^J - g_{IJ} F^I_{\mu \nu} F^{J|\mu \nu} + \cdots\;,
\] 
where it is understood that the scalars $h^I $ are constrained
to the hypersurface $h(h^I)=1$, and where the omitted terms
are not relevant for the solutions under consideration (i.e. the given
terms must correspond to a consistent truncation).

We now review the use of the local $r$-map in constructing
five-dimensional black hole solutions in the context 
of generalized special geometry, i.e. the Hesse potential $h$
is homogeneous of arbitrary degree $p$.
The relation between the five-dimensional and four-dimensional
metric is:
\begin{equation}
\label{Decomp5d}
ds_{(5)}^2 = - e^{2\tilde{\sigma}} (dt + {\cal A}_m dx^m)^2 
+ e^{-\tilde{\sigma}} ds_{(4)}^2 \;,
\end{equation}
where $\tilde{\sigma}$ is the Kaluza-Klein scalar and where 
${\cal A}_m$ is the Kaluza-Klein vector. The decomposition 
has been chosen such that the gravitational part of the 
action remains in the canonical Einstein Hilbert form 
upon reduction. 
We will focus on static (non-rotating) solutions, 
characterized by a vanishing Kaluza-Klein vector. We will
also impose that the four-dimensional Euclidean metric $ds_{(4)}^2$
is flat. We will refer to this class of solutions as extremal 
static solutions, because it includes extremal static black holes.

The reduced Lagrangian takes the form
\[
{\cal L} \sim - \frac{1}{2} R_{(4)} - g_{IJ} (\partial_m \sigma^I 
\partial^m \sigma^I - \partial_m b^I \partial^m b^J)+ \cdots \;.\
\]

To solve the Einstein equations with a flat metric, 
we need to impose that the 
energy momentum tensor of the four-dimensional theory vanishes. 
This can be achieved by imposing the extremal instanton ansatz
\[
\partial_m \sigma^I = \pm \partial_m b^I \;, 
\]
which selects totally isotropic submanifolds of $N$. 
For $p=3$, supersymmetric
solutions correspond to taking the same sign for all pairs of fields,
while different choices of signs correspond to non-supersymmetric 
extremal solutions. If the scalar metric $g_{IJ}$ has discrete isometries,
\[
g_{IJ} R^I_{\;\;K} R^J_{\;\;L} = g_{LM} \;,
\]
one can generalize the ansatz to
\[
\partial_m \sigma^I = R^I_{\;\;J} \partial_m b^J \;.
\]
This corresponds to `rotating the charges' in the corresponding
black hole solutions, which is a technique for obtaining 
non-supersymmetric solutions from supersymmetric solutions 
\cite{Ceresole:2007wx,LopesCardoso:2007ky}. 

It is clear that the totally isotropic
submanifolds picked by the ansatz are totally geodesic and 
hence lead to harmonic maps into $N$, because we are choosing
eigendirections of the para-complex structure. Since $N$ is
para-K\"ahler, the eigendistributions of the para-complex 
structure are not only integrable and totally geodesic, but parallel
and flat with respect to the Levi-Civita connection \cite{Cortes:2009cs,Mohaupt:2009iq}. It is possible
to verify this directly by using suitable local coordinates, as we 
will review now. 

In a flat background, the scalar equations of motion are
\begin{eqnarray}
\partial^m (g_{IJ} \partial_m \sigma^J) - \frac{1}{2}\partial_I
g_{JK}(\partial_m \sigma^J \partial^m \sigma^K - \partial_m b^J 
\partial^m b^K) &=& 0\;, \nonumber \\
\partial^m(g_{IJ} \partial_m b^J) &=& 0 \;. 
\end{eqnarray}
After imposing the extremal instanton ansatz or its 
generalization, this reduces to
\[
\partial^m (g_{IJ} \partial_m \sigma^J) = 0 \;.
\]
For the Hessian metric $g_{IJ}$ we can define dual coordinates
$\sigma_I$ by
\[
\partial_m \sigma_I = g_{IJ} \partial_m \sigma^J \;.
\]
Note that since $g_{IJ}$ is homogeneous of degree $-2$, this
implies $\sigma_I = - g_{IJ} \sigma^J$. In terms of dual coordinates,
the scalar equations reduce to decoupled harmonic equations
\[
\partial^m \partial_m \sigma_I = 0 \;,
\]
and the solution is given in terms of $n$ harmonic functions
$H_I$,
\begin{equation}
\label{5dsol}
\sigma_I = H_I \;.
\end{equation}
The choice 
\[
H_I = h_I + \sum_{k=1}^N \frac{q_{I,k}}{|x-x_{(k)}|^2}
\]
leads to multi-centered instanton solutions which lift 
to multi-centred extremal black hole solutions, with charges
$q_{I,k}$ located at the centers $x_{(k)}$, which correspond
to the positions of the event horizons. Expressing the 
solution (\ref{5dsol}) in terms of five-dimensional quantities,
namely the Kaluza-Klein scalar $\tilde{\sigma}$ and the 
constrained scalars $h^I$, we recover the five-dimensional
attractor equations \cite{Chamseddine:1998yv,Chamseddine:1999qs}
\[
e^{-\tilde{\sigma}} \frac{\partial h}{\partial h^I} = H_I \;.
\]
The line element of the black hole is determined by
\[
e^{\tilde{\sigma}} = h(\sigma)^{1/p} \;,
\]
where $p$ is the degree of the Hesse potential $h$.
The ADM mass is given by
\[
M_{ADM} = \frac{3}{2} \oint_{S^3_\infty} d^3 \Sigma^m e^{-\tilde{\sigma}}
\partial_m \tilde{\sigma} \;,
\]
and it can be shown that this agrees with the instanton action 
of the underlying four-dimensional Euclidean solution \cite{Mohaupt:2009iq}.\footnote{There
are some subtleties concerning the zero modes of the axionic scalars,
which are discussed in \cite{Mohaupt:2010du}.}
Taking the near horizon limit of the attractor equations, one
obtains the attractor equations
\[
Z \left. \frac{\partial h}{\partial h^I} \right|_* = q_I \;,
\]
where * denotes evaluation on the horizon, and where
\[
Z = (r^2 e^{-\tilde{\sigma}})_* \;.
\]
These equations determine the horizon values $h^I_*$ in terms of the
charges $q_I$, which is the celebrated black hole attractor mechanism.
For $p=3$, $Z$ agrees with the central charge of the supersymmetry 
algebra, up to normalization, $Z=\frac{1}{p} q_I h^I_*$. 

For illustration let us give the explicit solution for the
Hesse potential $h=\sigma^1 \ldots \sigma^p$ \cite{Mohaupt:2009iq}. 
For $p=3$ this corresponds
to the so-called STU-model.
The line element is
\[
ds_{(5)}^2 = - (H_1 \cdots H_p)^{-2/p} dt^2 +
(H_1 \cdots H_p)^{1/p} \delta_{mn} dx^m dx^n\;,
\]
and the five-dimensional scalars
\[
h^I = \left( \frac{ \prod_{K\not=I} H_K }{H_I^{p-1}} \right)^{1/p}
\]
have the limit
\[
h^I \rightarrow 
\left( \frac{ \prod_{K\not=I} q_{m,K}}{H_{m,I}^{p-1}} \right)^{1/p}
\]
at the $m$-th center. The entropy of the $m$-th center is
\[
{\cal S}_m= \frac{\pi^2}{2} Z^{3/2}_m = \frac{\pi^2}{2} 
\sqrt{q_{1,m} \ldots q_{p,m}} \;.
\]

\subsection{Generating solutions from the local $c$-map}

When adapting the above construction to the local $c$-map
one encounters additional complications. This is partly 
due to the fact that four-dimensional black holes can carry
both electric and magnetic charges. If one switches off the
magnetic charges, and discards supersymmetry, then the 
construction reviewed in the previous section is straightforward
to adapt to arbitrary dimensions, with only minor changes in 
the numerical values of coefficients (in particular the
decomposition of the metric (\ref{Decomp5d})).

However, the new form of the $c$-map based on the Hesse potential
and special real coordinates allows to generalize the
above construction to the case of the local $c$-map \cite{MohVau}. 
The question
whether there is a generalized version of special K\"ahler geometry
to which the construction can be extended remains open. The problem
is that any such generalization will introduce an asymmetry 
(different scaling weights) between
electric and magnetic degrees of freedom, and it is currently not
clear how to include this into the formalism. Therefore we remain
within the realm of special K\"ahler geometry proper.

Including time-like reductions in the previous calculations is 
straightforward and only leads to a few sign changes, which 
we parametrize by $\epsilon =-1$ for space-like and 
$\epsilon =1$ for time-like reduction. After dimensional reduction
one obtains the following three-dimensional Lagrangian for the bosonic
fields \cite{MohVau}
\begin{eqnarray}\label{3dLagrangian}
{\cal L} &\sim& - \frac{1}{2} R_3 - \tilde{H}_{ab}
(\partial_m q^a \partial^m q^b - \epsilon \partial_m \hat{q}^a
\partial_m \hat{q}^b) \\
&& - \frac{1}{H^2} (q^a \Omega_{ab} \partial_m q^b)^2 + \epsilon
\frac{2}{H^2} (q^a \Omega_{ab} \partial_m \hat{q}^b )^2 \nonumber \\
&& - \frac{1}{4H^2} (\partial_m\tilde{\phi} + 2 \hat{q}^a \Omega_{ab}
\partial_m \hat{q}^b ) \;. \nonumber
\end{eqnarray}
This is the Einstein-Hilbert term combined with a non-linear 
sigma model, and for $\epsilon=-1$ we recover the metric (\ref{MetricQK}).
We focus on solutions which will generate extremal static black holes
and impose that the three-dimensional metric is flat.\footnote{Note that
in three dimensions Ricci flatness already implies flatness. In higher 
dimensions we could generalize our construction to allow a Ricci flat
metric on the reduced space. This is left to future work.}
To solve the Einstein equations, the energy momentum tensor must
vanish identically, which selects totally isotropic submanifolds 
of $\bar{Q}$. For $\epsilon=1$ it is clear on general grounds that
$\bar{Q}$ is para-Quaternion-K\"ahler \cite{Cortes:2003zd,Cortes:2005uq}, 
but we will not use this
directly. Rather we will describe how the equations of motion 
can be solved explicitly in terms of harmonic functions, using
dual coordinates and suitable ans\"atze. Since the scalar Lagrangian
is a combination of perfect squares, a standard strategy is 
to impose the Bogomol'nyi equations resulting from the vanishing
of these squares. 

In particular, the term 
$\tilde{H}_{ab}
(\partial_m q^a \partial^m q^b - \partial_m \hat{q}^a
\partial_m \hat{q}^b)$ in the first line (setting $\epsilon =1$) 
is analogous to the case of the local $r$-map. This motivates us to impose
the extremal instanton ansatz
\[
\partial_m q^a = \pm \partial_m \hat{q}^a \;,
\]
and to introduce dual coordinates defined by
\[
\partial_m q_a = \tilde{H}_{ab} \partial_m q^b  \;.
\]
Using that $\tilde{H}_{ab}$ is homogeneous of degree $-2$, and
various other previous results, we obtain
\[
q_a = \tilde{H}_a= - \tilde{H}_{ab} q^b = \frac{1}{H} 
(-v_I, u^I) \;,
\]
where
\[
u^I \simeq \mbox{Im} X^I \;,\;\;\;v_I \simeq \mbox{Im} F_I
\]
are proportional the standard dual special real coordinates. More precisely, 
affine special K\"ahler manifolds do not only admit one
special connection, but a whole $S^1$-family thereof, which is
generated by the action of the complex structure \cite{Freed:1997dp,Alekseevsky:1999ts,Cortes:2009cs}. Whereas
$\mbox{Re}X^I$, $\mbox{Re}F_I$ are affine coordinates
for the original special connection, $\mbox{Im}X^I$ and $\mbox{Im}F_I$ 
are affine
coordinates for the `opposite' family member with parameter
value $\theta =\pi$, if we parametrize the $S^1$ familiy by
an angle $0\leq \theta < 2\pi$.
The variables $q_a$ and $u^I, v_I$ used above are related to the 
dual special real coordinates $\mbox{Im} X^I, \mbox{Im}F_I$ through
rescaling by specific powers of $H$, which itself is proportional to 
the Kaluza-Klein scalar $\phi$ \cite{MohVau}.

Compared to the previous section, we have additional terms
in the second and third line of (\ref{3dLagrangian}). These
terms are not independent once we impose the extremal instanton
ansatz. Due to the relative factor 2, the terms in the second line
do not cancel but combine into a single terms. The same term
appears within the perfect square in the third line. Now we
remark that the term in the third line is proportional to the
square of the field strength of the Kaluza-Klein vector. Thus
imposing that the third line vanishes implies that the resulting
solution is static (non-rotating). It is possible to obtain
more general, rotating solutions, by only requiring that 
the sum of second and third line vanishes. We refer to \cite{MohVau}
for this case and focus on static solutions, where the second and
third line vanish separately. 

It is then straightforward, though somewhat tedious to verify
that the scalar equations reduce to 
\[
\partial^m \partial_m q_a = 0 \;,
\]
while all other equations are satisfied identically. Thus, as 
before, the solution is given in terms of harmonic functions.

By rewriting the equations
\[
q_a = H_a \;,
\]
where $H_a$ are $2n+2$ harmonic functions, in terms of the 
four-dimensional variables, we recover the four-dimensional
black hole attractor equations\footnote{When comparing to 
\cite{LopesCardoso:2000qm}, note that $e^{2f}=\phi^{-1}$, where 
$f$ is the function used to parametrize the four-dimensional
metric in \cite{LopesCardoso:2000qm}. Also note that the variables
called $Y^I$ in \cite{LopesCardoso:2000qm} differ from the $Y^I$ used
in this article by a factor $\phi$. Also, when comparing to 
\cite{MohVau}, remember to replace $\phi \rightarrow e^\phi$.}
\[
\phi^{-1/2} (X^I - \bar{X}^I) = i H^I \;,\;\;\;
\phi^{-1/2} (F_I - \bar{F}_I) = i H_I \;,
\]
in a symplectically covariant form.

Moreover
when rewriting the extremal instanton ansatz 
\[
\partial_m q^a = \pm \partial_m \hat{q}^a
\]
in terms of four-dimensional quantities we obtain
\[
\partial_m (\phi^{1/2} (X^I + \bar{X}^I )) = \mp 
(F^{I|+}_{0m} + F^{I|-}_{0m}) \;,\;\;\;
\partial_m (\phi^{1/2} (F_I + \bar{F}_I )) = 
\mp (G^+_{I|0m} + G^-_{I|0m})) \;,\;\;\;
\]
which shows that the real part of the 
symplectic vector $(X^I,F_I)$ is proportional to the electric and 
magnetic potentials. 
For supersymmetric solutions this follows from
the gaugino variation \cite{Behrndt:1997ny,LopesCardoso:2000qm}, while 
here we obtain it as the 
the Bogomol'nyi equation associated to
the first line of (\ref{3dLagrangian}).

Not surprisingly our formalism shows many similarities with the
superconformal approach and its emphasis on symplectic covariance.
Since we work with the bosonic field equations rather than with 
Killing spinor equations, we can also obtain non-supersymmetric
extremal solutions. Supersymmetric solutions have the same 
sign for all fields in the instanton ansatz. Non-supersymmetric 
solutions correspond to the generalized ansatz, 
\[
\partial_m q^a = R^a_{\;\;b} \partial_m \hat{q}^b
\]
where the 
matrix $R^a_{\;\;b}$ is a discrete isometry of $\tilde{H}_{ab}$.
We refer to \cite{Mohaupt:2010fk,MohVau} for
results on non-supersymmetric, non-extremal
and rotating solutions. Note that non-extremal black hole 
solutions have also been
discussed recently in \cite{Galli:2011fq,Meessen:2011bd}.

\ack This article is based on a talk 
given by T.M. at the QTS7 in Prague.
He thanks the organizers of QTS7 for 
the invitation and for organising 
this inspiring conference.
The work of T.M. is supported in part by STFC  
grants ST/G00062X/1 and ST/J000493/1.
The work of O.V. is supported by an STFC studentship and by DAAD. 
Part of the material included in this article is based
on unpublished work by V. Cort\'es and the first author. 
Both authors thank V. Cort\'es for useful discussions and
ongoing collaboration on the topics reviewed in this article.
T.M. also thanks B. de Wit, J. Louis, P. Smyth 
and S. Vandoren for useful discussions.

\providecommand{\newblock}{}

\end{document}